# Level statistics of deformed even-even nuclei

A. Al-Sayed and A. Y. Abul-Magd
*Faculty of Science, Zagazig University, Zagazig, Egypt*

Abstract
The nearest neighbor spacing distribution of levels of deformed even-even nuclei classified according to their quadrupole deformation parameter is investigated. The results suggest that the oblate deformed nuclei have more regular spectra than prolate ones.

## 1. Introduction

The interest of chaotic motion and non-linear dynamics in atomic nucleus has risen sharply in last two decades. Level statistics provide a major tool to investigate such chaotic behavior within nucleus. Wigner [1] introduced random matrix theory (RMT) to model the energy levels of nuclei excited at neutron threshold energies. Bohigas et al. [2] conjectured that the Gaussian Orthogonal Ensemble (GOE) provides a satisfactory description for the fluctuation properties of generic quantum systems with time-reversal symmetry, which in the classical limit are fully chaotic. Level spacing distributions of neutron and proton s-wave resonances agree with GOE predictions. The question of whether the chaotic behavior applies to nuclear dynamics in the ground state region was revised in ref. [3]. The nearest neighbor spacing (NNS) distribution at low lying excitation energies requires complete (few or no missing levels) and pure (few or no unknown spin parities) level schemes. Unfortunately, complete and pure level schemes are only available for a limited number of nuclei. So to perform a significant statistical study, we in need to combine different sequences of levels (each sequence has the same spin parity) from different nuclides. Such statistical studies showed that, the NNS distributions are intermediate between Poisson distribution for purely ordered systems, and Wigner distribution for purely chaotic systems [4,5,6]. In addition to this, light nuclei behave close to chaotic systems, while heavier nuclei seem to be more regular.

In the present study, we investigate the effect of oblate- and prolate-shape on the chaoticity of deformed even-even nuclei. We focus on the $2^+$ states for their relative abundance in even-even nuclei. The data set and the deformation parameter used to classify the deformed nuclei is described in sec. 2. Section 3 describes the Bayesian technique used in the analysis, while our results are given in section 4.

## 2. Data Set

In this section, we describe our choice of levels in deformed nuclei. We recall that the collective motion is interpreted as vibrations and rotations of the nuclear surface in the geometric collective model first proposed by Bohr and Mottelson [7], where a nucleus modeled as a charged liquid drop. The moving nuclear surface may be described quite generally by an expansion in spherical harmonics with time-dependent shape parameters as coefficients [8]:

*E-mail addresses:* abdallahalsayed@gmail.com (A. Al-Sayed)
a_y_abul_magd@hotmail.com (A.Y. Abul-Magd)

$$R(\theta,\phi,t) = R_{av}[1+\sum_{\lambda=0}^{\infty}\sum_{\mu=-\lambda}^{\lambda}\alpha_{\lambda\mu}(t)Y_{\lambda\mu}(\theta,\phi)], \quad (1)$$

where $R(\theta,\phi,t)$ denotes the nuclear radius in the direction $(\theta,\phi)$ at time t, $R_{av}$ is the radius of the spherical nucleus. For $\lambda$=0 terms correspond to change in nuclear radius (need a large amount of energy for the compression of nuclear matter). The $\lambda$=1 terms correspond to a translation of the centre-of-mass so; we normally excluded them from the sum. The $\lambda$=2, the quadrupole deformation seems to be the most important collective excitations of the nucleus. $\lambda$ and $\mu$ determine the surface co-ordinates as functions of $\theta$ *and* $\phi$ respectively. For axially symmetric nuclei, we can rewrite the nuclear radius as,

$$R(\theta,\phi) = R_{av}[1+\beta_2 Y_{20}(\theta,\phi)], \quad (2)$$

the quadrupole deformation parameter $\beta_2$ (=$\alpha_{20}$), can be related to the axes of the spheroid by [9],

$$\beta_2 = \frac{4}{3}\sqrt{\frac{\pi}{5}}\frac{\Delta R}{R_{av}}, \quad (3)$$

in which the average radius, $R_{av}=R_0A^{1/3}$, and $\Delta R$ is the difference between the semi-major and semi-minor axes. The larger the value of $\beta_2$ the more deformed the nucleus. Positive and negative $\beta_2$ values correspond to prolate and oblate shapes respectively.

We qualify nuclei as deformed according to the liquid drop model calculation by P. Moller et al. [10]. We thus ignore spherical nuclei for which, the deformation parameter $\beta_2$ is equal to zero.

The data on low-lying $2^+$ levels of deformed even-even nuclei are taken from Endt [11] for $22 \leq A \leq 44$, and from the Nuclear Data Sheets up to March 2005 for heavier nuclei. We consider nuclei in which the spin-parity $J^\pi$ assignments of at least five consecutive levels are definite. In cases where the spin-parity assignments are uncertain and where the most probable value appeared in brackets, we accept this value. We terminate the sequence in each nucleus when we arrive at a level with unassigned $J^\pi$, or when an ambiguous assignment involved a spin-parity among several possibilities, as e.g. $J^\pi = (2^+, 4^+)$. We made an exception when only one such level occurred and was followed by several definitely assigned levels containing at least two levels of the same spin-parity, provided that this ambiguous level is found in a similar position in the spectrum of a neighboring nucleus. However, this situation has occurred for less than 5% of the levels considered. A detailed description of the data set used in the present analysis is given in table (1). In this way, we obtained 30 nuclides of oblate deformation having 246 energy levels, and 84 nuclides of prolate deformation having 595 energy levels.

## 3. Method of Analysis

In this section, we perform a statistical analysis for the NNS distribution of prolate and oblate nuclei separately. For the requirement of statistical studies using RMT, we should have a spectrum of unit mean level spacing. This is done by fitting a theoretical expression to the number *N(E)* of levels below excitation energy *E,* this process is called unfolding. The expression used here is the constant-temperature formula (4),

$$N(E) = N_0 + \exp\left(\frac{E - E_0}{T}\right). \quad (4)$$

The three parameters $N_0$, $E_0$ and $T$ obtained for each nucleus vary considerably with mass number. Nevertheless, all three show a clear tendency to decrease with increasing mass number. We fitted all nuclei as a function of mass number $A$ using 2nd order polynomial function for each the three unfolding parameters. We got the following values for each fitting parameters,

$T = (2.2\pm0.2) - (0.017\pm0.003)\,A + (4\pm1)\text{x}10^{-5}\,A^2$,

$E_0 = (4.2\pm0.5) - (0.044\pm0.008)\,A + (1.3\ \pm3)\text{x}10^{-5}\,A^2$,

$N_0 = (0.7\pm0.4) - (0.005\pm0.007)\,A + (2\ \pm3)\text{x}10^{-5}\,A^2$.

We got these values after excluding five nuclei from all nuclei under investigation (114 nuclei) where they are semimagic or at subshell closure [12] at $N$ or $Z$ equal to 38. Where the level density near closed shells are smaller than elsewhere. These nuclei are $^{72}$Se, $^{206}$Pb, $^{82}$Sr, $^{92}$Sr, and $^{96}$Mo (excluded where $N$=54). A detailed account of this method has been given in ref. [13, 14], and references within.

The nuclear states are characterized by their invariance under time reversal and space rotation, which could be represented by the Gaussian orthogonal ensemble (GOE) of random matrices. The NNS distribution of levels of the GOE is well approximated by Wigner's surmise [1]

$$Pw(s) = \frac{\pi}{2} s \exp\left(-\frac{\pi}{4} s^2\right). \quad (5)$$

Here, $s$ is the spacing of neighboring levels in units of the mean level spacing. For integrable systems, the NNS distribution is generically given by the Poisson distribution,

$$P_P(s) = \exp(-s). \quad (6)$$

Here we confine ourselves to the central aspects. The key ingredient of this analysis is the assumption that the deviation of the NNS distribution of low-lying nuclear levels from the GOE statistics is caused by the neglect of possibly existing conserved quantum numbers other than energy, spin, and parity. A given sequence $S$ of levels can then be represented as a superposition of $m$ independent sequences $S_j$ each having fractional level density $f_j$, with $j = 1, \ldots, m,$ and with $0 < f_j \le 1$ and $\sum_{j=1}^{m} f_j = 1$. We assume that the NNS distribution $P_j(s)$ of $S_j$ obeys GOE statistics. The exact NNS distribution $P(s)$ has been given in ref. [15]. It depends on the $(m-1)$ parameters $f_j$, $j = 1, \ldots, m-1$. In [16], this expression has been simplified by observing that $P(s)$ is mainly determined by short-range level correlations. This reduces the number of parameters to unity and the proposed NNS distribution of the spectrum is

$$P(s, f) = \left[1 - f + f(0.7 + 0.3f)\frac{\pi s}{2}\right] \times \exp\left(-(1-f)s - f(0.7+0.3f)\frac{\pi s^2}{4}\right), \quad (7)$$

which depends on only a single parameter, the mean fractional level number $f = \sum_{j=1}^{m} f_j^2$ for the superimposed sub-spectra. This quantity will eventually be used as a fit parameter. For a large number $m$ of sub-spectra, $f$ is of the order of $1/m$. In this limit, $P(s, f)$ approaches the Poisson distribution $P(s, 0) = P_P(s)$. This expresses the well-known fact that the superposition of many GOE level-sequences produces a Poissonian sequence. On the other hand, when $f \to 1$ the spectrum approaches the GOE behavior. Indeed, $P(s, 1)$ coincides with the Wigner distribution (5) which is expected as the system in this case consists of a single GOE sequence. This is why $f$ is refered as the chaoticity parameter.

The Bayesian inference method is used to determine the parameter $f$, the Bayesian analysis consists in determining the best-fit value of the chaoticity parameter $f$ and its error for each NNS distribution. When $P(f|s)$ is not Gaussian, the best-fit value of $f$ cannot be taken as the most probable value. Rather we take the best-fit value to be the mean value $\bar{f}$ and measure the error by the standard deviation $\sigma$ of the posterior distribution, i.e.

$$\bar{f} = \int_0^1 f P(f\,|\,s)\,\mathrm{d}f\,, \text{ and } \sigma^2 = \int_0^1 (f - \bar{f})^2 P(f\,|\,s)\,\mathrm{d}f\,. \qquad (8)$$

The result of the analysis is shown in fig. (1). We firstly note that, Eq. (7) provides a satisfactory description of the NNS distribution. The best-fit values of the parameter are $f = 0.72 \pm 0.04$ for prolate nuclei, and $f = 0.59 \pm 0.07$ for oblate ones.

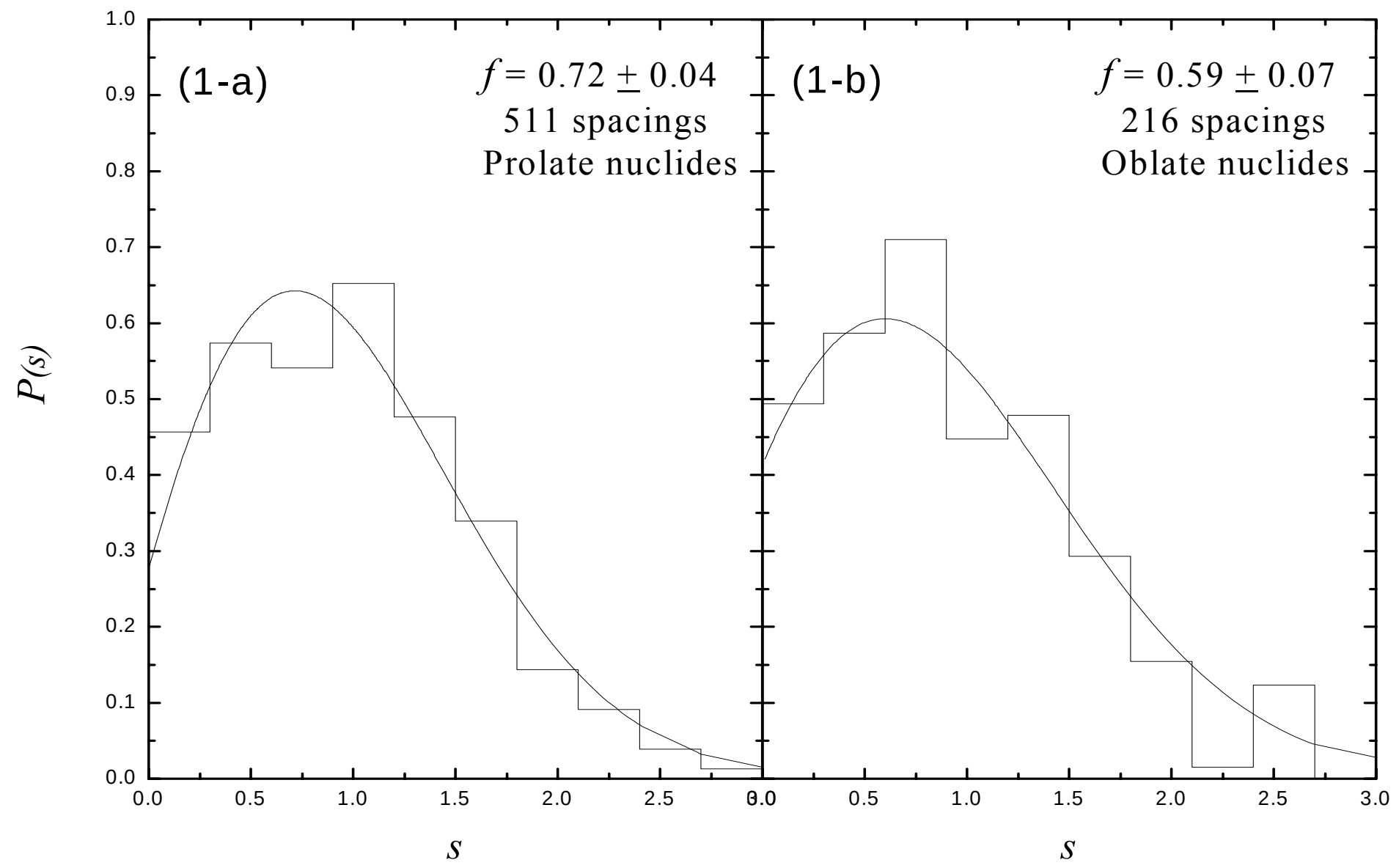

Fig. (1): The chaoticity of prolate- oblate deformed even-even nuclei.

## 4. Summary and Conclusion

In the present paper, we study the nearest neighbor spacing distribution of deformed even-even nuclei using the available experimental energy levels. We use a model that interpolates between Poisson (regular) to Wigner (chaotic) distribution by varying the chaoticity parameter from 0 to 1 respectively. The difference in the chaoticity parameter of each of the deformation groups is statistically significant. Our analysis shows that the chaoticity parameter is smaller in the oblate deformed nuclei than the prolate ones, suggesting that the former group is more regular than the latter. It is interesting to note that the work of Arvieu et al. [17], and Mukhopathyay and Pal [18] suggested that, the single particle motion in a deformed potential is more chaotic in case of oblate deformation than in prolate case. However, our model assumes that the intrinsic motion of nuclei is chaotic. The apparent regularity of the spectrum is because of the approximate conservation of the quantum number describes the collective degrees of freedom. Our results may be interpreted that the coupling between the single particle and collective degrees of freedom is weaker in oblate than prolate nuclei.

## 5. References

[1] E. P. Wigner, Oak Ridge National Laboratory Report No. ORNL-2309, (1957).

[2] O. Bohigas, M. J. Giannoni, and C. Schmit, Phys. Rev. Lett. 52, 1 (1984).

[3] A. Y. Abul-Magd and H. A. Weidenmüller, Phys. Lett. B 162, 223 (1985).

[4] T. von Egidy, A.N. Behkami, H.H. Schmidt, Nucl. Phys. A 454, 109 (1986).

[5] T. von Egidy, H. H. Schmidt and A. N. Behkami, Nucl. Phys. A 481,189 (1988).

[6] J.F. Shriner, G.E. Mitchell and T. von Egidy, Z. Phys. A338, 309 (1991).

[7] A. Bohr and B. R. Mottelson, Nuclear structure, Vol.II, Nuclear Deformations, Benjamin, New York, (1975).

[8] W. Greiner and J. A. Maruhn, Nuclear Models, Springer-Verlag, Berlin Heidelberg (1996).

[9] C. Wheldon, Ph.D. thesis, "K-isomerism at high-spin beyond the fusion limit", http://www.hmi.de/people/wheldon/thesis/thesis/node10.html .

[10] P. Moller, J. R. Nix, W. D. Myers, and W. J. Swiatecki, Atomic and Nuclear Data Tables, 59, 185 (1995).

[11] P.M. Endt, Nucl. Phys. A 633, 1 (1998).

[12] R.F. Casten and N.V. Zamfir, J. Phys. G, Vol. 22, 1521 (1996).

[13] A.Y. Abul-Magd, et al., Phys. Lett. B 579, 278 (2004).

[14] A.Y. Abul-Magd, et al., Ann. of Phys., Vol. 321, 560 (2006).

[15] M. L. Mehta, Random Matrices, 2nd Ed., Academic, New York, (1991).

[16] A. Y. Abul-Magd and M. H. Simbel, Phys. Rev. E 54, 3292 (1996); Phys. Rev. C 54, 1675 (1996).

[17] R. Arvieu, F. Brut, J. Carbonell, and J. Touchard, Phys. Rev. A 35, 2389 (1987).

[18] T. Mukhopathyay and S. Pal, Nucl. Phys. A 592, 291 (1995).

| Oblate Nuclei | | | Prolate Nuclei | | | | | | | | |
|---|---|---|---|---|---|---|---|---|---|---|---|
| Nucleus | n | $E_{max}$ | Nucleus | n | $E_{max}$ | Nucleus | n | $E_{max}$ | Nucleus | n | $E_{max}$ |
| $^{28}$Si | 33 | 14.5774 | $^{112}$Sn | 10 | 3.2864 | $^{114}$Cd | 10 | 2.6501 | $^{154}$Gd | 5 | 1.2942 |
| $^{26}$Mg | 13 | 8.8638 | $^{66}$Ni | 5 | 3.746 | $^{150}$Gd | 7 | 2.1799 | $^{164}$Yb | 7 | 1.513 |
| $^{72}$Se | 7 | 3.2135 | $^{90}$Zr | 7 | 5.2752 | $^{192}$Os | 11 | 1.8679 | $^{182}$W | 5 | 1.8569 |
| $^{68}$Ge | 5 | 3.0231 | $^{88}$Sr | 10 | 4.414 | $^{228}$Ra | 5 | 1.0132 | $^{180}$Pt | 5 | 1.3512 |
| $^{74}$Se | 6 | 2.482 | $^{70}$Zn | 6 | 2.665 | $^{58}$Fe | 14 | 4.62 | $^{182}$Pt | 6 | 1.3109 |
| $^{116}$Cd | 7 | 2.29247 | $^{94}$Mo | 5 | 2.7398 | $^{108}$Pd | 9 | 2.47757 | $^{250}$Cf | 7 | 1.4113 |
| $^{70}$Ge | 7 | 3.3346 | $^{96}$Ru | 6 | 2.65 | $^{146}$Ba | 6 | 1.9684 | $^{160}$Dy | 8 | 1.8699 |
| $^{76}$Se | 7 | 3.017 | $^{82}$Sr | 5 | 2.885 | $^{148}$Nd | 5 | 1.6599 | $^{162}$Er | 7 | 1.8649 |
| $^{72}$Ge | 5 | 3.0942 | $^{84}$Sr | 5 | 3.175 | $^{150}$Sm | 10 | 2.0545 | $^{166}$Yb | 5 | 1.5798 |
| $^{74}$Ge | 8 | 3.017 | $^{92}$Zr | 10 | 3.26262 | $^{152}$Gd | 10 | 1.9757 | $^{106}$Ru | 5 | 1.8856 |
| $^{66}$Zn | 9 | 3.6707 | $^{84}$Kr | 5 | 2.775 | 154Dy | 5 | 1.5077 | $^{162}$Dy | 8 | 1.9991 |
| $^{188}$Pt | 5 | 1.5281 | $^{88}$Kr | 6 | 2.6511 | $^{160}$Yb | 7 | 1.8113 | $^{164}$Dy | 5 | 1.7381 |
| $^{190}$Pt | 7 | 1.6007 | $^{82}$Kr | 8 | 3.4575 | $^{188}$Os | 7 | 1.8076 | $^{168}$Er | 10 | 2.1932 |
| $^{192}$Pt | 9 | 1.5764 | $^{96}$Mo | 6 | 2.7869 | $^{228}$Th | 8 | 1.1754 | $^{172}$Hf | 6 | 1.4823 |
| $^{194}$Pt | 8 | 1.87 | $^{92}$Sr | 5 | 2.8209 | $^{230}$Th | 6 | 1.4009 | $^{174}$Hf | 5 | 1.4964 |
| $^{140}$Sm | 5 | 2.2899 | $^{200}$Pt | 5 | 1.842 | $^{232}$Th | 7 | 1.3872 | $^{178}$Hf | 9 | 1.8913 |
| $^{196}$Pt | 9 | 1.8537 | $^{136}$Ce | 6 | 2.6838 | $^{232}$U | 5 | 1.1328 | $^{170}$Yb | 5 | 1.5346 |
| $^{198}$Pt | 5 | 1.6369 | $^{106}$Cd | 8 | 2.7179 | $^{60}$Fe | 8 | 4.176 | $^{170}$Er | 6 | 1.4163 |
| $^{192}$Hg | 6 | 2.0817 | $^{108}$Cd | 8 | 2.6826 | $^{62}$Zn | 8 | 3.83 | $^{172}$Yb | 5 | 1.6085 |
| $^{196}$Hg | 9 | 1.979 | $^{132}$Ba | 5 | 2.0463 | $^{64}$Zn | 5 | 3.0713 | $^{28}$Mg | 5 | 5.6738 |
| $^{198}$Hg | 9 | 1.9015 | $^{76}$Ge | 6 | 2.5911 | $^{96}$Zr | 10 | 3.6022 | $^{76}$Kr | 7 | 2.697 |
| $^{200}$Hg | 8 | 1.9723 | $^{110}$Cd | 20 | 2.9753 | $^{110}$Pd | 9 | 2.1401 | $^{22}$Ne | 10 | 8.596 |
| $^{124}$Te | 5 | 2.323 | $^{112}$Cd | 6 | 2.2311 | $^{122}$Xe | 5 | 2.0654 | $^{100}$Zr | 5 | 1.8076 |
| $^{62}$Ni | 12 | 3.9729 | $^{144}$Ce | 9 | 2.4052 | $^{156}$Dy | 6 | 1.515 | $^{24}$Mg | 8 | 10.361 |
| $^{202}$Hg | 5 | 1.3897 | $^{54}$Cr | 6 | 3.8611 | $^{234}$U | 6 | 1.1742 | | | |
| $^{200}$Pt | 5 | 1.842 | $^{80}$Se | 5 | 2.344 | $^{240}$Pu | 5 | 1.1805 | | | |
| $^{204}$Hg | 8 | 2.1174 | $^{82}$Se | 9 | 4.391 | $^{246}$Cm | 7 | 1.7808 | | | |
| $^{204}$Pb | 9 | 2.3163 | $^{98}$Mo | 10 | 2.4183 | $^{184}$W | 9 | 1.8767 | | | |
| $^{206}$Pb | 7 | 2.984 | $^{100}$Ru | 9 | 2.3514 | $^{100}$Mo | 5 | 2.0428 | | | |
| $^{214}$Po | 8 | 2.0108 | $^{106}$Pd | 9 | 2.5004 | $^{152}$Sm | 6 | 1.2928 | | | |

Table (1): A description of available data, symbol *n* represents the number of $2^+$ states, and $E_{max}$ for the highest $2^+$ level contributed for each nucleus. The first three columns are devoted to oblate deformed nuclei, while the rest columns for prolate ones.